# Growth and band alignment of $Bi_2Se_3$ topological insulator on H-terminated Si(111) van der Waals surface

Handong Li, Lei Gao, Hui Li, Gaoyun Wang, Jiang Wu, Zhihua Zhou, and Zhiming Wang[a]

*State Key Laboratory of Electronic Thin Films and Integrated Devices, University of Electronic Science and Technology of China, Chengdu 610054, China*

**ABSTRACT**

The van der Waals epitaxy of single crystalline $Bi_2Se_3$ film was achieved on hydrogen passivated Si(111) (H:Si) substrate by physical vapor deposition. Valence band structures of $Bi_2Se_3$/H:Si heterojunction were investigated by X-ray Photoemission Spectroscopy and Ultraviolet Photoemission Spectroscopy. The measured Schottky barrier height at the $Bi_2Se_3$-H:Si interface was 0.31 eV. The findings pave the way for economically preparing heterojunctions and multilayers of layered compound families of topological insulators.

PACS: 81.15.Kk, 68.55.-a, 82.80.Pv, 73.30.+y

As a prototype topological insulator (TI), $Bi_2Se_3$ is characterized with a large bulk band gap (~ 0.3 eV) intercrossed by single-cone-shaped gapless surface states.[1,2] The single Dirac cone of $Bi_2Se_3$ is expected to facilitate the exploring of novel quantum phenomenon in TI, and the wide energy band gap will promise a great possibility for room temperature application. Among the efforts in exploring the transport behaviors of $Bi_2Se_3$, one featured progress has been made by Mclver JW *et al.* that spin-to-momentum conversion (SMC) on the Dirac cone surface states can be detected.[3] It also shows that the SMC can be observed in $Bi_2Se_3$ sample with carrier density as high as ~ $1 \times 10^{19}$ $cm^{-3}$, although a high background doping level of TIs has been thought to hinder access to surface chiral carriers.[4-6] However, as an analogue to spin-galvanic effect,[7] the necessary non-equilibrium spin population to drive the orthogonal-orientated charge current was optically pumped in the experiment of Ref. 3. More recently, a full-electrical SMC scheme was demonstrated by introducing a ferromagnetic metal (FM) spin injector for $Bi_2Se_3$ TI rather than employing optical orientation procedure used in Ref. 3.[8] To isolate the stray field of FM electrode from $Bi_2Se_3$, a Si spacer layer has to be inserted between FM and TI. Since spin polarized carriers can be effectively injected from the FM electrode and drift quite a long distance through Si without loss of phase coherence,[9] the SMC efficiency in the proposed device will thus subject to the spin injection efficiency at the $Bi_2Se_3$-Si heterointerface.

However, critical information for understanding spin transport behaviors of the $Bi_2Se_3$/Si heterojunction such as the relative band positions is still lacking. In this work, thin film growth strategy was employed in order to prepare $Bi_2Se_3$/Si junction. X-ray Photoemission Spectroscopy (XPS) and Ultraviolet Photoemission Spectroscopy (UPS) were conducted to study the core levels (CLs) and valence bands of both thick and thin $Bi_2Se_3$ films on Si, respectively. We will show that high quality $Bi_2Se_3$ films can be grown on H-terminated Si(111) (H:Si hereafter) substrates by economical physical vapor deposition (PVD) method. The measured CLs show no chemical shifts between thick (85nm) and thin (10nm) samples further supports the van der Waals epitaxy (vdWE) [10-13] nature of $Bi_2Se_3$ on H:Si. Valence band study indicates the Fermi level of as-grown $Bi_2Se_3$ lies in the conduction band and the calculated Fermi level position is 0.37 eV higher than the $Bi_2Se_3$ conduction band minimum. Based on XPS and UPS measurements, Shottky barrier height of $Bi_2Se_3$/H:Si junction is calculated to be 0.31 eV. The band alignment at the $Bi_2Se_3$/H:Si interface is also pictured.

The growth of the $Bi_2Se_3$ film was carried out in a home-built PVD facility. A base pressure of 0.5 Pa could be obtained by using a set of rotary pump. High purity (>99.999%) $Bi_2Se_3$ pellets (Alfa Aesar Inc.) were used as source materials and placed at the hot center of the furnace during the deposition. The resistivity of the Si(111) substrate used in this study was ~ 100 $\Omega \cdot cm$. H-terminated van der Waals surface of Si(111) was produced by dipping clean Si(111) substrate into diluted HF solution for 10 minute. Prior to the etching process, standard SPM process was employed for cleaning the Si(111) substrates. High purity Argon (>99.999%) was used as feeding gas and the flow rate was held at 0.3 SLM by a mass flow control during the growth process. As soon as the hot center of the furnace was heated to growth temperature (550°C), the Ar flow was supplied which triggered the thin film deposition process. During the growth, the H:Si substrates were in temperature range of 200-250°C and the film thickness could be adjusted by changing growth time. After growth, the films were naturally cooled down to room temperature then taken out and cut into 5 × 5 mm$^2$ samples for further characterizations.

Before performing XPS/UPS studies, structural and surface morphology details of the as-grown films were characterized by X-ray diffraction (XRD) and scanning electron microscopy (SEM), respectively. Using Ecopia HMS-2000 Hall system, carrier concentrations and mobilities of Bi$_2$Se$_3$ films were measured. The sample thickness was examined by a Veeco Dekta 150 profilometer. The XPS and UPS measurements (from Omicron GmbH) were carried out *ex situ* in a UHV chamber (~ 5 × 10$^{-10}$ mbar) which was equipped with a monochromatized X-ray Al Kα source (hν1486.7 eV) for XPS, and a He I source (21.2 eV) for UPS, respectively. The probe area during the XPS and UPS measurements was adjusted to ~ 1 mm$^2$ by using the imaging facilities of the system.

The growth of Bi$_2$Se$_3$ on H:Si by PVD, in spite of the 7.8% lattice mismatch, exhibited epitaxial nature as confirmed by structural characterizations. Typical XRD θ-2θ scan of a 110 nm thick Bi$_2$Se$_3$ film grown on H:Si substrate was shown in Figure 1 (a). Only the (00n) (n = 3, 6, 9, 12, …) diffraction peaks of Bi$_2$Se$_3$ could be observed indicated the c-axis preferred orientation of the film. In typical XRD Φ scan of Bi$_2$Se$_3$ (015) for a thinner Bi$_2$Se$_3$ film (~ 30 nm), six broad peaks were observed [figure 1(b)] in the scan range of 0°–360° which depicted in-plane hexagonal ordering of the film. In-plane rotating of Bi$_2$Se$_3$ epifilm with respect to the H:Si substrate was negligible as could be seen from the Φ scan peaks of Si(220) [figure 1(c)]. Since rhombohedral crystal lattice of Bi$_2$Se$_3$ exhibits three-fold symmetry around the [001] axis, twinning defects were expected to reside in the epifilm. Accordingly, two overlapping epitaxial relationships characterized respectively as Bi$_2$Se$_3$(001)[1 1 0]//Si(111)[$\bar{1}$ $\bar{1}$ 2] and Bi$_2$Se$_3$(001)[1 $\bar{1}$ 0]//Si(111)[112] were inferred to coexist in the epitaxial film.

Figure 2 (a) was an SEM image showing the surface features near a macroscopic scratch on Bi$_2$Se$_3$ film. Large size Bi$_2$Se$_3$ slabs with smooth surface could be observed. In a magnified SEM image, the growth front features of PVD-grown Bi$_2$Se$_3$ film could be clearly illustrated. As could be seen in figure 2 (b), straight bunched steps rather than facets which originated from surface trianglar spirals (indicated by dashed triangles) and ran across the whole sample surface were clearly figured out. These surface spiral domains were 180° twinned, which were similar with those morphologies observed in the Bi$_2$Se$_3$ films grown by molecular beam epitaxy (MBE) approaches.[14-17] Actually, the unique surface characterizations manifested a spiral growth mode of the PVD-grown film which was also quite identical to the MBE-grown cases. The Bi$_2$Se$_3$ film grown by PVD was n-type and highly conductive as measured by van der Pauw Hall with silver paste cured at room temperature used for the contacts. Typical carrier concentration and mobility values were $5 \times 10^{18}$ cm$^{-3}$ and 800 cm$^2$V$^{-1}$S$^{-1}$, respectively, which were also comparable to that of Bi$_2$Se$_3$ films grown on Si by MBE.[14]

In the following, XPS and UPS spectrum of Bi$_2$Se$_3$ on H:Si were studied. To access the CLs' signals from Bi$_2$Se$_3$/H:Si heterointerface, a thick film (110 nm) was firstly thinned by a 3 keV Ar ion sputtering process after it had been loaded into the XPS/UPS analysis chamber. The etch rate of Bi$_2$Se$_3$ film during sputtering was ~ 5 nm/min as calibrated by another Bi$_2$Se$_3$ film sample and the etching was uniform in depth all over the whole sample area as manifested by thickness measurements after sputtering for several minutes. As shown in Figure 3 were the XPS spectrum recorded during the etching process of a thick $Bi_2Se_3$ film. No sample charging effect was observed due to the high conductivity of $Bi_2Se_3$ sample. Curve (a) was from the as-grown surface of the sample, on which C1s and O1s could be clearly seen. After sputtering for 10 min, the two peaks vanished and the rest peaks could be attributed exactly to Bi and Se which indicated a completely removal of contamination on the sample surface as curve (b) shown. The Si 2*p* and 2*s* peaks began to appear at a nominal thickness of 15 nm as indicated in curve (c). As thickness decreasing, the intensity of Si peaks grew rapidly as shown in curve (d)-(f) which indicated that the CLs of both substrate and film were quite accessible at thickness lower than 15 nm. To monitor the surface structure evolution of the $Bi_2Se_3$ film during the sputtering, low electron energy diffraction (LEED) was employed and no diffraction spots from Si could be observed even as nominal film thickness was down to 4 nm. It indicated unambiguously not only a uniform covering of $Bi_2Se_3$ thin film on H:Si substrate but also was the thickness uniformity of $Bi_2Se_3$ grown by PVD method better than ± 4nm. Both the chemical shifts and the normalized peak intensity proportion of Bi and Se kept constant further depicted the compositional stability of $Bi_2Se_3$ surface during the Ar ion sputtering process.

After removal of surface contamination, the $Bi_2Se_3$ film was inspected by UPS and XPS, respectively. Figure 4(a) presented the UPS valence band spectrum from a cleaned $Bi_2Se_3$ surface (85 nm). In the binding energy range from -2 eV to 6 eV, the UPS curve of the $Bi_2Se_3$ exhibits two well defined regions. One was at low binding energy where a peak located at around ~ 2.1 eV which could be attributed to Bi 6*p* antibonding.[18,19] While at E = 0 eV, a small peak could be clearly seen which indicated significant electron occupation in the conduction band. Indeed, since the carrier density of $Bi_2Se_3$ film was shown to be as high as ~ $10^{18}$ cm$^{-3}$, Fermi level was expected to locate in the conduction band. The VBM of Bi$_2$Se$_3$ ($E_{\text{Bi}_{\text{VBM}}}^{\text{Bi}_2\text{Se}_3}$) was determined by the intersection between linear fitting to the leading edge of the spectrum and the background[20] and a value of 0.67 eV was obtained. Therefore the Fermi level position in the conduction band was 0.37 eV higher than the conduction band minimum if the band gap of Bi$_2$Se$_3$ was taken as 0.3 eV.

The Bi 5$d$ core level spectrum of both thick (85nm) and thin (10nm) Bi$_2$Se$_3$ film were measured by XPS and shown in Figure 4(b) and (c), respectively. Being a typical heavy metal, spin-orbit splitting in Bi was so large that peaks of 5$d_{3/2}$, 5$d_{5/2}$ core levels were clearly separated. However, to precisely determine the peaks' positions, Lorentz–Gauss profiles and Shirley background had been taken for the deconvolution. The binding energies of Bi 5$d_{5/2}$ from thick ($E_{\text{Bi}5d_{5/2}}^{\text{Bi}_2\text{Se}_3}$) and thin films ($E_{\text{Bi}5d_{5/2}}^{\text{Bi}_2\text{Se}_3}(i)$) were determined to be 25.27 eV and 25.22 eV respectively. Such a small chemical shift indicated no interfacial reaction occurred between Bi$_2$Se$_3$ and H:Si which unambiguously evidenced the vdWE nature of Bi$_2$Se$_3$ grown on H:Si.

The Si 2$p$ CLs from the thin Bi$_2$Se$_3$ film (10nm) covered H:Si were also studied. For the Si 2$p$ spectrum, curve fitting schemes were employed to distinguish the precise positions of overlapped 2$p_{3/2}$ and 2$p_{1/2}$ peaks. The curves were constrained by documented spin-orbit splitting,[21] intensity ratio, and FWHM ratio of H:Si peak given in this work. The best fit based on the given constraints was shown in figure 4(d) and Si 2$p_{3/2}$ position, $E_{\text{Si}2p_{3/2}}^{\text{H:Si}}(i)$ was 99.70 eV.

Using the values of CLs and VBM obtained in XPS and UPS measurements, the valence band offset (VBO) can thus be calculated by using the Kraut equation:[22]

$$E_{\text{VBO}} = E_{\text{CL}} + (E_{\text{Si}2p_{3/2}}^{\text{H:Si}} - E_{\text{Si}_{\text{VBM}}}^{\text{H:Si}}) - (E_{\text{Bi}5d_{5/2}}^{\text{Bi}_2\text{Se}_3} - E_{\text{Bi}_{\text{VBM}}}^{\text{Bi}_2\text{Se}_3}), \quad (1)$$

and the Schottky barrier height $\Phi_\text{B}$ is given by

$$\Phi_\text{B} = (E_{\text{Si}2p_{3/2}}^{\text{H:Si}} - E_{\text{Si}_{\text{VBM}}}^{\text{H:Si}}) + E_\text{G}^{\text{H:Si}} - E_{\text{Si}2p_{3/2}}^{\text{H:Si}}(i), \quad (2)$$

Where $E_{\text{CL}} = E_{\text{Bi}5d_{5/2}}^{\text{Bi}_2\text{Se}_3}(i) - E_{\text{Si}2p_{3/2}}^{\text{H:Si}}(i)$ is the CL offset at the interface. $E_\text{G}^{\text{H:Si}}$, $E_{\text{Si}2p_{3/2}}^{\text{H:Si}}$, and $E_{\text{Si}_{\text{VBM}}}^{\text{H:Si}}$ are values of band gap, CL energy, and VBM from bulk H:Si, respectively.

From Eq. (1), the $E_{\text{VBO}}$, can be calculated with measured values of $(E_{\text{Bi}5d_{5/2}}^{\text{Bi}_2\text{Se}_3} - E_{\text{Bi}_{\text{VBM}}}^{\text{Bi}_2\text{Se}_3})$, $E_{\text{CL}}$, and documented value of $(E_{\text{Si}2p_{3/2}}^{\text{H:Si}} - E_{\text{Si}_{\text{VBM}}}^{\text{H:Si}})$[23], and is 0.19 eV. From Eq. (2), Schottky barrier height $\Phi_\text{B}$ can be calculated with measured value of $E_{\text{Si}2p_{3/2}}^{\text{H:Si}}(i)$ and documented values of $E_\text{G}^{\text{H:Si}}$ and $(E_{\text{Si}2p_{3/2}}^{\text{H:Si}} - E_{\text{Si}_{\text{VBM}}}^{\text{H:Si}})$, and is 0.31 eV. Compared with the electronically measured Schottky barrier (0.34 eV),[8] the XPS/UPS measured value in our work is smaller. This may be due to the silicon substrates employed in our experiment have higher resistivity (i.e., higher electron affinity). Moreover, the sharp interface between the epitaxial $Bi_2Se_3$ layer and H:Si surface will contribute to eliminating interfacial potential fluctuations thus an ideal Schottky junction structure is expected. Finally, the band diagram of $Bi_2Se_3$/H:Si heterojunction is schematically indicated in figure 5.

In summary, we successfully prepared single crystalline $Bi_2Se_3$ films on H-terminated Si(111) surface by PVD method. Based on precise XPS and UPS measurements, the Schottky barrier and band alignment of $Bi_2Se_3$/H:Si heterostructure were calculated and illustrated, respectively. The vdWE scheme, which had only been considered responsible for MBE growth of layered compounds on chemically inert substrates with large lattice mismatch, was proved valid for the more economical PVD method. Another advantage of growing $Bi_2Se_3$ on H:Si van der Waals surface is that the TI Dirac cone of $Bi_2Se_3$ at the interface can be effectively preserved. Indeed, very small chemical shifts (0.05 eV) of Bi $5d$ CLs between bulk $Bi_2Se_3$ and $Bi_2Se_3$/H:Si interface was experimentally confirmed in this work which reflected the weak bonding nature between $Bi_2Se_3$ epilayers and H:Si substrates. Considering the layered compound families of TIs share similar crystal structure of $Bi_2Se_3$, our findings may serve as useful reference for preparing heterostructures and multilayers of them in an economical way.

**ACKNOWLEDGEMENTS**

This work is supported by the National Natural Science Foundation of China under Grant No. 11104010 and the Fundamental Research Funds for the Central Universities of China under Grant No. ZYGX2012J033.

**FIGURE LEGENDS**

FIG. 1. (a) XRD out-of-plane θ-2θ scan of a 110 nm thick $Bi_2Se_3$ thin film grown on H:Si substrate. The in-plane Φ scans of $Bi_2Se_3$ (015) and of Si (220) from the same sample were shown in (b) and (c), respectively. The thickness of the film used in XRD Φ scans was 30 nm.

FIG. 2. (a) Large-scale SEM image shows the surface morphology near a scratch on a 110 nm thick $Bi_2Se_3$ film. (b) A magnified SEM image taken on the same sample illustrates the step bunching features of $Bi_2Se_3$ surface. The dashed triangles pointed left and right respectively indicate the existence of rotation domain pairs.

FIG. 3. (Color online) XPS measurements on $Bi_2Se_3$ film on H:Si(111) before (a) and after Ar ion sputtering (b)-(e). The nominal film thickness of $Bi_2Se_3$ in (b) to (e) is 85 nm, 15 nm, 10 nm, 6 nm, and 4 nm, respectively. Inset is a typical LEED image (E = 75 eV) taken from 10 nm $Bi_2Se_3$ film during the sputtering.

FIG. 4. (Color online) (a) Valence band maximum measured by UPS and (b) Core level energy spectrum measured by XPS for cleaned $Bi_2Se_3$ film with nominal thichness of 85nm. Core level energy spectrum measured by XPS for $Bi_2Se_3$ film with nominal thickness of 10 nm on H:Si(111) are shown in (c) and (d). Dotted curves are original data.

FIG. 5. (Color online) The schematic band diagram for $Bi_2Se_3$ on H:Si(111) heterojunction system. $E_C^{Bi_2Se_3}$ and $E_C^{H:Si}$ represent conduction band minima of $Bi_2Se_3$ and H:Si, respectively.

**FIGURES**

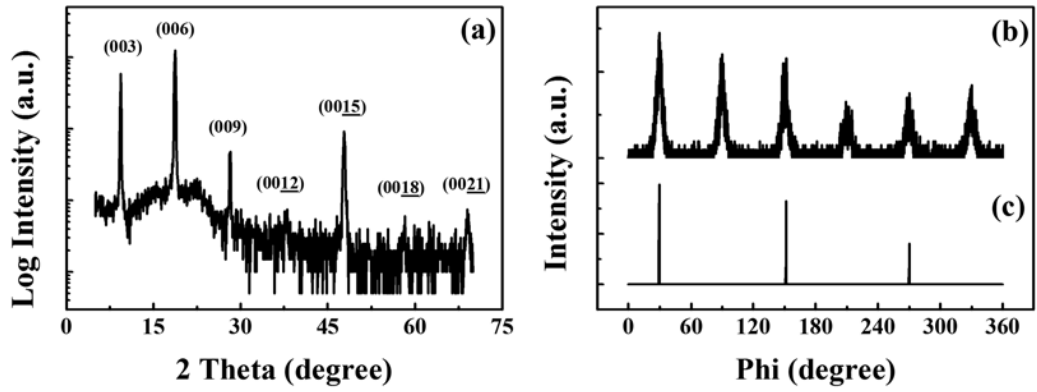

FIG. 1.

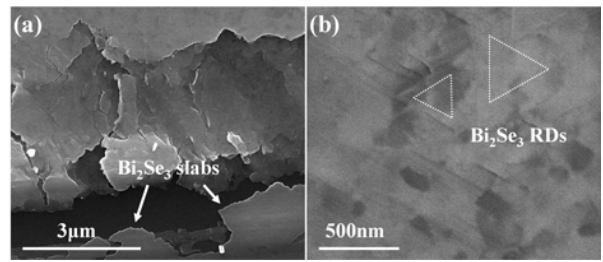

FIG. 2.

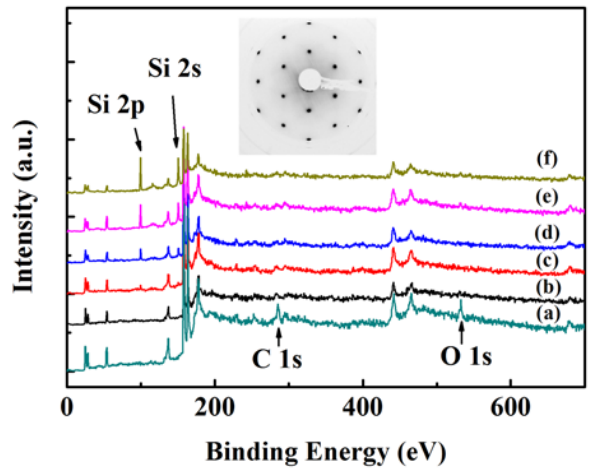

FIG. 3.

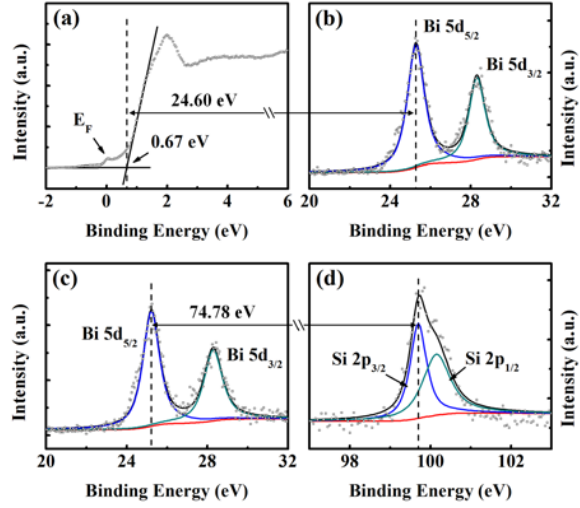

FIG. 4.

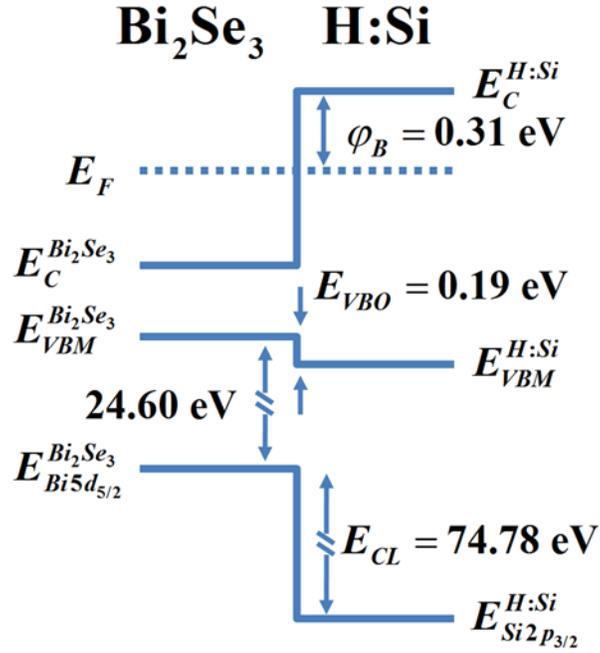

FIG. 5.